\documentclass[12pt, preprint]{aastex} \usepackage{epsf}
\newcommand{\Lx}{L_{\mathrm{X}}}
\shorttitle{Mass-temperature relation of galaxy clusters}
\shortauthors{Shimizu et al.}
\begin{document}
\received{2002 August 30} \accepted{2002 February 24}

\title{Mass-temperature relation of galaxy clusters: implications
  from the observed luminosity-temperature relation and X-ray
  temperature function}

\author{\sc Mamoru Shimizu, \altaffilmark{1} Tetsu Kitayama,
  \altaffilmark{2}  Shin Sasaki\altaffilmark{3}, and Yasushi
  Suto\altaffilmark{1}}

\email{mshimizu@utap.phys.s.u-tokyo.ac.jp,
  kitayama@ph.sci.toho-u.ac.jp,  sasaki@phys.metro-u.ac.jp,
  suto@phys.s.u-tokyo.ac.jp}

\altaffiltext{1}{Department of Physics,  School of Science,  The
  University of Tokyo,  Tokyo 113-0033,  Japan}

\altaffiltext{2}{Department of Physics,  Toho University,  Funabashi,
  Chiba 274-8510,  Japan}

\altaffiltext{3}{Department of Physics, Tokyo Metropolitan University,
  Hachioji, Tokyo 192-0397, Japan}


\begin{abstract}
  We derive constraints on the mass-temperature relation of galaxy
  clusters from their observed luminosity-temperature relation and
  X-ray temperature function. Adopting the isothermal gas in
  hydrostatic equilibrium embedded in the universal density profile of
  dark matter halos, we compute the X-ray luminosity for clusters as a
  function of their hosting halo mass. We find that in order to
  reproduce the two observational statistics, the mass-temperature
  relation is fairly well constrained as $T_{\rm gas} =(1.5\sim 2.0)\; {\rm
    keV} (M_{\rm vir}/10^{14}h_{70}^{-1}M_\odot)^{0.5\sim 0.55}$, and a simple
  self-similar evolution model ($T_{\rm gas} \propto M_{\rm
    vir}^{2/3}$) is strongly disfavored. In the cosmological model
  that we assume (a $\Lambda$CDM universe with $\Omega_0=0.3$,
  $\lambda_0=0.7$ and $h_{70}=1$), the derived mass-temperature
  relation suggests that the mass fluctuation amplitude $\sigma_8$ is
  0.7--0.8.
\end{abstract}

\section{Introduction}

While clusters of galaxies are relatively simple dynamical systems
that consist of dark matter, stars, and X-ray--emitting hot gas, their
thermal evolution is not yet fully understood. This is clearly
illustrated by the well-known inconsistency of the observed X-ray
luminosity-temperature ($\Lx$-$T$) relation,  $\Lx \propto T^3$
\citep[e.g., ][]{david93, markevitch98, ae99} against the simple
self-similar prediction $\Lx \propto T^2$ \citep{kaiser86}.
Conventionally this is interpreted as evidence for preheating of
intracluster gas;  additional heating tends to increase the
temperature and the core size of the cluster, and to decrease the
central density and the luminosity. Since the effect is stronger for
less massive systems,  the slope of $\Lx$-$T$ relation becomes steeper
than that of the self-similar prediction \citep{eh91, kaiser91}.

In addition, the mass-temperature ($M$-$T$) relation of clusters is
also poorly determined. Although it is conventionally assumed that the
gas shock heating is efficient enough and the temperature of the
intracluster gas reaches the corresponding virial temperature of the
hosting halos, this should be regarded as a simple working hypothesis.
Nevertheless, the cosmological parameters derived from the cluster
abundances are sensitive to the adopted $M$-$T$ relation. While this
has already been recognized for some time (e.g., Figs.5d and 6d of
Kitayama \& Suto 1997), \citet{seljak02} recently showed in a
quantitative manner that the use of the observed $M$-$T$ relation by
\citet*{finogu} decreases the value of the mass fluctuation amplitude
at $8\; h^{-1}\;$Mpc,  $\sigma_8$, by $\sim 20\%$ where $h$ is the
Hubble constant $H_0$ in units of 100 km s$^{-1}$ Mpc$^{-1}$ [we use,
however, the dimensionless Hubble constant $h_{70} \equiv
H_{0}/(70\;\mathrm{km}\; \mathrm{s}^{-1}\; \mathrm{Mpc}^{-1})$ in the
following analysis]. Therefore, the independent derivation of the
cluster $M$-$T$ relation is important both in understanding the
thermal history of the intracluster gas and in determining the
cosmological parameters.

Our primary aim in this paper is to find the $M$-$T$ relation of
clusters that reproduces the observed $\Lx$-$T$ relation and X-ray
temperature function (XTF). The reason we focus on $M$-$T$ relation is
as follows:  since recent $N$-body simulations strongly indicate the
universality of the density profile of the hosting halos of clusters
(\citealt*{nfw96}; \citealt{moore98, js00}), the intracluster gas
density profile in hydrostatic equilibrium with the underlying dark
matter can be computed \citep*{MSS98, SSM98} for a given mass of the
halo. This enables one to make a reliable prediction for the $\Lx$-$T$
relation once the $M$-$T$ relation is specified. In turn, one can
obtain the $M$-$T$ relation that reproduces the observed $\Lx$-$T$
relation without assuming an ad hoc model for the thermal evolution of
intracluster gas.

In what follows, we parameterize the $M$-$T$ relation as a single
power law, and derive the best-fit values of their amplitude and slope
from the observed $\Lx$-$T$ relation and the XTF. The result is
compared with the recent observational studies by \citet{finogu} and
\citet*{allen}. We also discuss the implications for the value of
$\sigma_8$ from cluster abundances. Throughout the paper, we adopt a
conventional $\Lambda$CDM model with density parameter
$\Omega_{0}=0.3$, cosmological constant $\lambda_{0}=0.7$,
dimensionless Hubble constant $h_{70}=1$,  and baryon density
parameter $\Omega_{{B}}=0.04\;h_{70}^{-2}$.

\section{Model of intracluster gas in dark matter halos}
\label{sec:densprof}

We first outline the model of the dark matter halo and the isothermal gas
density profile embedded in the halo, which are essential in predicting
the X-ray luminosity of clusters as a function of the mass of the
hosting halo.

\subsection{Dark Matter Density Profile}

We adopt the specific density profile of dark matter halos of mass
$M_{\rm vir}$, given as
\begin{equation}
 \label{eq:nfw}
 \rho_{\rm halo}(r; M_{\rm vir})=
 \left\{
  \begin{array}{cc}
   \displaystyle
   \frac{\bar{\rho}(z) \, \delta_{c}(M_{\rm vir})}{
    (r/r_{s})^{\alpha} (1+r/r_{s})^{3-\alpha}} 
   & r< r_{\rm vir}, \\
   \displaystyle
   0 & r>r_{\rm vir},
  \end{array}
 \right.
\end{equation}
where $\bar\rho(z) \equiv \Omega_0 \rho_{c0} (1+z)^3$ is the mean
density of the universe at $z$, $\rho_{c0}$ is the present
critical density, $\delta_{c}(M_{\rm vir})$ is the
characteristic density excess, and $r_{\rm vir}$ and $r_{s}$
are the virial radius and the scale radius of the halo, respectively.
In practice, we focus on two specific profiles:  $\alpha=1$
\citep{nfw96} and $\alpha=3/2$, indicated by higher resolution
simulations \citep{moore98, js00, fukushigemakino01}.

The virial radius $r_{\rm vir}$ is defined according to the spherical
collapse model as
\begin{equation}
 r_{\rm vir}(M_{\rm vir})
 \equiv \left(\frac{3M_{\rm vir}}{4\pi\bar{\rho} \Delta_{\rm nl}}
  \right)^{1/3}, 
 \label{eq: r_vir}
\end{equation}
and the approximation for the critical overdensity $\Delta_{\rm nl}=
\Delta_{\rm nl}(\Omega_0, \lambda_0)$ can be found in \citet{KS96}.
The two parameters $r_{s}$ and $r_{\rm vir}$ are related via the
concentration parameter,
\begin{equation}
 c=c(M_{\rm vir}, z)
 \equiv
 \frac{r_{\rm vir}(M_{\rm vir}, z)}
 {r_{s}(M_{\rm vir}, z)}.  
 \label{eq: concentration}
\end{equation}

In the case of $\alpha=1$, we use an approximate fitting function with
the same functional form as that of \citet{Bullock01},
\begin{equation}
 c_{{B}}(M_{\rm vir}, z)
 =\frac{c_{\mathrm{norm}}}{1+z}\; \left(
  \frac{M_{\rm vir}}{1.4\times 10^{14}\; h_{70}^{-1}\; M_{\odot}}
 \right)^{-0.13}. 
 \label{eq:c_Bullock}
\end{equation}
Since the $c$-$M_{\rm vir}$ relation has a fairly large intrinsic scatter,
we adopt the same value of the power-law index ($-0.13$) as indicated by
\citet{Bullock01}, but perform our own fit to their Figure 4 to
determine the value of the coefficient $c_{\rm norm}$.  In fact, the
original coefficient given by \citet{Bullock01} does not seem to fit
their data, and we find $c_{\mathrm{norm}}=8 {+ 2~~ \atop - 2.7}$. The
quoted errors do not represent the fitting error to the mean relation,
but correspond to $\pm 1\;\sigma$ for the intrinsic distribution around
the mean relation.  The uncertainty with respect to the adopted
power-law index ($-0.13$) is effectively included in the above quoted
errors for $c_{\rm norm}$.  For $\alpha \not=1$, we rescale the
amplitude of the concentration parameter according to \citet{keeton01}
as $c(M_{\rm vir}, z)=(2-\alpha)c_{{B}}(M_{\rm vir}, z)$.

The condition that the total mass inside $r_{\rm vir}$ is equal to
$M_{\rm vir}$ relates $\delta_{c}$ to $c$ as
\begin{equation}
 \delta_{c}(M_{\rm vir})
 = {\Delta_{\rm nl} \over 3}\frac{c^{3}}{m(c)}, 
\end{equation}
where
\begin{equation}
\label{eq:nfw-m}
 m(x)=\left\{
 \begin{array}{cc}
  \displaystyle
  {\ln(1+x)-x/(1+x)} & \alpha=1,\\
  \displaystyle
  2\left[{\ln(\sqrt{x}+\sqrt{1+x})-\sqrt{\frac{x}{1+x}}}\right]
   & \alpha=3/2.
  \end{array}
 \right.
\end{equation}

\subsection{Hydrostatic Equilibrium Gas Distribution}
\label{sec:gasdensprof}

We further assume that the intracluster gas is isothermal and in
hydrostatic equilibrium, which is a reasonable physical approximation.
Under the gravitational potential of the above dark matter halos, the
isothermal gas density profiles in hydrostatic equilibrium are
computed analytically as  \citep{SSM98}
\begin{equation}
 \rho_{\rm gas}(r)=\rho_{\mathrm{gas}, 0}\; \exp[-Bf(r/r_{{s}})], 
 \label{eq:rhogas}
\end{equation}
where
\begin{equation}
\label{eq:b}
 B=\frac{2c}{m(c)}\frac{T_{\rm vir}}{T_{\rm gas}}
\end{equation}
and
\begin{equation}
 f(x)= \left\{
  \begin{array}{cc}
   \displaystyle
   1-\frac{1}{x}\ln (1+x)& \alpha=1,\\
   \displaystyle
   2\sqrt{\frac{1+x}{x}}-\frac{2}{x}\ln(\sqrt{x}+\sqrt{1+x})&
   \alpha=3/2.
  \end{array}
 \right.
\end{equation}
In equation (\ref{eq:b}), $T_{\rm vir}$ is the virial temperature,
which we define as
\begin{equation}
 k_{\mathrm{B}}T_{\rm vir}=\frac{1}{2}\mu m_{{p}}
 \frac{GM_{\rm vir}}{r_{\rm vir}} \propto M_{\rm vir}^{2/3}, 
 \label{eq:tvir}
\end{equation}
where $k_{\mathrm{B}}$ is the Boltzmann constant, $G$ is the
gravitational constant, and $\mu m_{{p}}$ is the mean molecular
weight. We adopt $\mu=0.6$ assuming that the gas is almost fully
ionized with the mass fractions of helium $Y=0.24$ and metals $Z=0.3\;
Z_\odot$ ($Z_\odot=0.02$). On the other hand, the gas temperature
$T_{\rm gas}$ is determined from the $M$-$T$ relation, as described in
detail in \S\ref{sec:masstemp}. The central gas density $\rho_{\rm
  gas, 0}$ is computed so that
\begin{equation}
\int_0^{r_{\rm vir}} \rho_{\rm gas}(r) 4 \pi r^2 dr = f_{\rm gas}M_{\rm vir} 
\left(\frac{\Omega_{B}}{\Omega_0}\right), 
\end{equation}
where $f_{\rm gas}$ is the hot gas fraction of baryon mass in the
cluster described in \S\ref{sec:fhot}. The X-ray luminosity of
clusters is computed as
\begin{equation}
\label{eq:lx}
 L_{\mathrm{X}}=4\pi\int^{r_{\rm vir}}_{0}\Lambda
(T_{\rm gas}, Z)
 \left[
 \frac{\rho_{\rm gas}(r)}{\mu m_{{p}}}
 \right]^{2}r^{2}dr.
\end{equation}
In practice, we adopt the bolometric cooling function of
\citet{sutherlanddopita} for $\Lambda(T_{\rm gas}, Z)$.

\subsection{Mass-Temperature Relation}
\label{sec:masstemp}

As emphasized in the above, we are mainly interested in the $M$-$T$
relation of clusters. Most of previous theoretical studies adopted the
self-similar relation for the $M$-$T$ relation: $T_{\rm gas} = T_{\rm
  vir} \propto M_{\rm vir}^{2/3}$. Recent observations, however,
indicate a departure from this relation. \citet{finogu}, for instance,
obtained
\begin{equation}
 T_{\rm ew}= (2.63\pm 0.07)\; \mathrm{keV}
 \left(
  \frac{M_{500}}{10^{14}\; h_{70}^{-1}\; M_{\odot}}
 \right)^{0.54\pm 0.02}, 
 \label{eq:finogu}
\end{equation}
where $M_{\Delta_{c}}$ is the total mass enclosed within the
radius $r_{\Delta_{c}}$ at which the mean interior density is
$\Delta_{c}$ times the critical density of the universe, and
$T_{\mathrm{ew}}$ is the emission-weighted temperature.
\citet{finogu} estimated $M_{500}$ from the observed X-ray luminosity
density profile assuming that the gas is in hydrostatic equilibrium.
Similarly \citet{allen} found
\begin{equation}
 T_{\mathrm{2500}}= (3.38\pm 0.42)\; \mathrm{keV}
 \left(
  \frac{M_{2500}}{10^{14}\; h_{70}^{-1}\;M_{\odot}}
 \right)^{0.65\pm 0.09}, 
 \label{eq:allen}
\end{equation}
where $T_{2500}$ is the gas-mass--weighted temperature within
$r_{2500}$ \citep*[see also][]{edgm02}.

In order to generalize the various choices in the $M$-$T$ relation, we
adopt the parameterization
\begin{equation}
 \label{eq:mtparam}
 T_{\rm gas}(M_{\rm vir})=T_{\mathrm{gas, 0}}
 \left(
  \frac{M_{\rm vir}}{10^{14}\; h_{70}^{-1}\;M_{\odot}}
 \right)^{p_{\mathit{MT}}}. 
\end{equation}
For reference, the self-similar model $T_{\rm gas}=T_{\rm vir}$ with
equation~(\ref{eq:tvir}) corresponds to $(T_{\mathrm{gas, 0}},
p_{\mathit{MT}}) = (1.1~{\rm keV}, \frac{2}{3})$. Since
$M_{500}$ and $M_{2500}$ quoted in equations~(\ref{eq:finogu}) and
(\ref{eq:allen}) are different from $M_{\rm vir}$, we translate them
to $M_{\rm vir}$, properly taking into account the dark matter density
profiles as described in Appendix~\ref{sec:massconvert} (the relations
between $M_{\Delta_{c}}$ and $M_{\rm vir}$ for several different
values for the overdensity $\Delta_{c}$ are plotted in
Fig.~\ref{fig:f09} below). In doing so, we also convert the cluster data
at the individual redshifts into those at $z=0$, taking into account the
difference of the Hubble parameter at $z$ and $H_0$ for our assumed
cosmology ($\Omega_0=0.3$, $\lambda_{0}=0.7$, and $h_{70}=1.0$). Then
we find that the observed samples are well fitted to the
relations
\begin{equation}
 T_{\rm gas}= \left\{
  \begin{array}{cc}
   \displaystyle
   (1.92\pm 0.06)\; \mathrm{keV}
   \left(
    \frac{M_{\rm vir}}{10^{14}\; h_{70}^{-1}\; M_{\odot}}
   \right)^{0.54\pm 0.02} & \alpha=1,\\
   \displaystyle
   (1.88\pm 0.06)\; \mathrm{keV}
   \left(
    \frac{M_{\rm vir}}{10^{14}\; h_{70}^{-1}\; M_{\odot}}
   \right)^{0.54\pm 0.02} & \alpha=3/2,\\
  \end{array}
 \right.
 \label{eq:mtobsfitf}
\end{equation}
for \citet{finogu}, and
\begin{equation}
 T_{\rm gas}= \left\{
  \begin{array}{cc}
   \displaystyle
   (1.53\pm 0.56)\; \mathrm{keV}
   \left(
    \frac{M_{\rm vir}}{10^{14}\; h_{70}^{-1}\; M_{\odot}}
   \right)^{0.57\pm 0.12} & \alpha=1,\\
   \displaystyle
   (1.45\pm 0.54)\; \mathrm{keV}
   \left(
    \frac{M_{\rm vir}}{10^{14}\; h_{70}^{-1}\; M_{\odot}}
   \right)^{0.59\pm 0.12} & \alpha=3/2,\\
  \end{array}
 \right.
 \label{eq:mtobsfita}
\end{equation}
for \citet{allen}.

\subsection{Hot Gas Mass Fraction}
\label{sec:fhot}

Hot gas fraction $f_{\rm gas}$ also plays a central role in predicting
the X-ray luminosity of clusters. In many theoretical analyses, it is
often assumed that $f_{\rm gas}$ is independent of the mass of the
hosting halos just for simplicity. Of course, this is not a good
approximation because the gas in less massive systems is expected to
have cooled more efficiently at high redshifts on average, and thus
$f_{\rm gas}$ should be a monotonically increasing function of the
halo mass. This qualitative feature is supported by both observations
and hydrodynamical simulations. \citet*{Mohr99}, for instance, measured
the gas mass fraction within $r_{500}$ as a function of gas
temperature in the range $3\; \mathrm{keV}\lesssim T_{\rm gas}\lesssim
10\; \mathrm{keV}$. Extrapolating their result to lower temperatures,
and assuming for simplicity that the gas mass fraction at the virial
radius is equal to that at $r_{500}$, we obtain
\begin{equation}
 \label{eq:fhotobs500}
 f_{\rm gas} = \mathrm{min}\left[0.92 h_{\rm 70}^{-3/2}
 \left(
  \frac{T_{\rm gas}}{6\; \mathrm{keV}}
 \right)^{0.34}, \ 1\right].
\end{equation}
This is our fiducial model for the hot gas fraction in the present
analysis, and for comparison, we also consider a simple model $f_{\rm
  gas}=0.8$, in which the gas mass fraction is independent of halo
mass and gas temperature.

Strictly speaking, we have to convert the value of the gas fraction of
\citet{Mohr99} defined at $r_{500}$ to that at the virial radius. In
practice, however, the observational error of the value is
significantly larger than the difference of the conversion, and thus
we assume here that $f_{\rm gas}(r_{500})=f_{\rm gas}(r_{\rm vir})$.
We put an additional condition that the hot gas to baryon fraction in
clusters inside their virial radius does not exceed the cosmological
baryon fraction for our assumed cosmological parameters
($\Omega_0=0.3$ and $\Omega_{B}=0.04\;h_{70}^{-2}$). In fact, the
observed gas to dark matter fraction has a large scatter (see Fig.~14
of Mohr et al. 1999) and is consistent with the upper bound that we
set here. Actually, if we adopt the observational $M$-$T$ relation of
\citet{finogu}, we can constrain the gas mass fraction so as to
reproduce the observed $\Lx$-$T$ relation and XTF. As shown in
Appendix B, the resulting constraint is fairly consistent with
equation~(\ref{eq:fhotobs500}).

\section{Mass-temperature relation}

\subsection{Constraints from the Luminosity-Temperature Relation}

As mentioned above, the simple self-similar model prediction $\Lx
\propto T^2$ is too shallow to be consistent with the observation
($\Lx \propto T^3$). This means that heating/cooling processes in
addition to the shock heating are important in the thermal evolution
of intracluster gas. Apart from the physical mechanism of the
additional thermal processes, there are three possibilities that might
modify the mass dependence of X-ray luminosity (see eq.~[\ref{eq:lx}])
and steepen the resulting $\Lx$-$T$ relation. First, the gas density
profile may be significantly flatter for less massive systems.
Second, the mass dependence of the hot gas mass fraction is strong as
$f_{\rm gas} \propto M_{\rm vir}^{1/3}$. Finally, the mass-temperature
relation is $T_{\rm gas} \propto M_{\rm vir}^{2/5}$. In practice, a
realistic model should be a combination of those three effects to some
extent. The gas density profile can be specified completely from the
hydrostatic equilibrium assumption. As for the hot gas mass fraction,
we adopt the observed relation as well as the simple constant
fraction. Then the model described in the previous section enables one
to compute the X-ray luminosity of a given halo, and to examine
whether the observed $\Lx$-$T$ relation can be reproduced from
observed $M$-$T$ relation and the gas mass fraction.

\begin{figure}[ht]
  \epsscale{0.8} \plotone{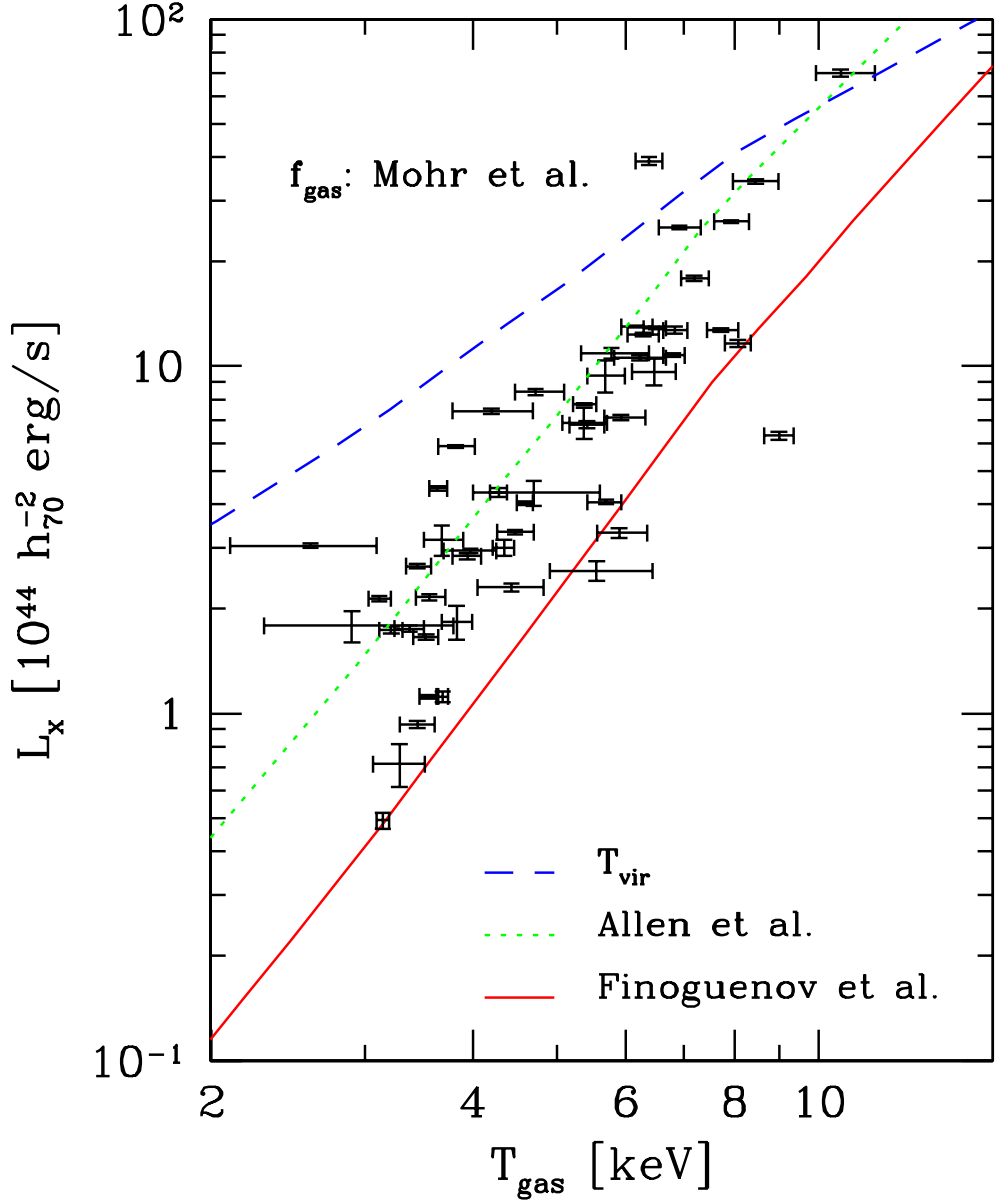}  \figcaption{X-ray $\Lx$-$T$
    relations derived from the observed $M$-$T$ relations
    (\textit{solid line}, eq.~[\ref{eq:mtobsfitf}]; \textit{dotted line},
    eq.~[\ref{eq:mtobsfita}]) assuming the
    observed gas mass fraction (eq.~[\ref{eq:fhotobs500}]). A simple
    self-similar model prediction with $T_{\rm gas}=T_{\rm vir}$ is
    plotted by the dashed line for comparison. The data with
    error bars indicate 52 X-ray clusters with temperature higher than
    2.5~keV from the sample of \cite{ikebe}. \label{fig:f01}}
\end{figure}

Before doing so, let us first look at the $\Lx$-$T$ relation
\textit{predicted} from the observed $M$-$T$ relations
(eqs.~[\ref{eq:mtobsfitf}] and [\ref{eq:mtobsfita}]) and gas mass
fraction (eq.~[\ref{eq:fhotobs500}]) combined with the isothermal gas
density profile (eq.~[\ref{eq:rhogas}]). Figure~\ref{fig:f01} compares
those predictions against 52 X-ray clusters with temperature higher
than 2.5~keV from the sample of \cite{ikebe} (excluding the two
clusters with no reliable estimates for their temperatures). For
comparison, we also plot the result for the self-similar model
(eq.~[\ref{eq:tvir}]). Incidentally, we performed all the analysis
below both for $\alpha=1$ and $3/2$, but their difference turns out to
be very small. Thus, we show the results for $\alpha=1$ except for the
combined contour plot (see Fig.~\ref{fig:f08} below).

This clearly illustrates that the predicted $\Lx$-$T$ relation is very
sensitive to the assumed $M$-$T$ relation; as is well known, the
simple self-similar model (Fig.~\ref{fig:f01}, \textit{dashed line})
is inconsistent with the observations by a wide margin. If the halo
density profile is well described by equation (\ref{eq:nfw}) with $1
\le \alpha \le 3/2$, the $M$-$T$ relation of \citeauthor{allen}
(\citeyear{allen}; Fig.~\ref{fig:f01}, \textit{dotted line}) is in
good agreement and that of \citeauthor{finogu} (\citeyear{finogu};
\textit{solid line}) leads to an acceptable result. This conclusion
in turn indicates that the $\Lx$-$T$ relation provides a good
diagnosis of the underlying $M$-$T$ relation, which is as yet poorly
determined observationally.

\begin{figure}[ht]
  \epsscale{0.8} \plotone{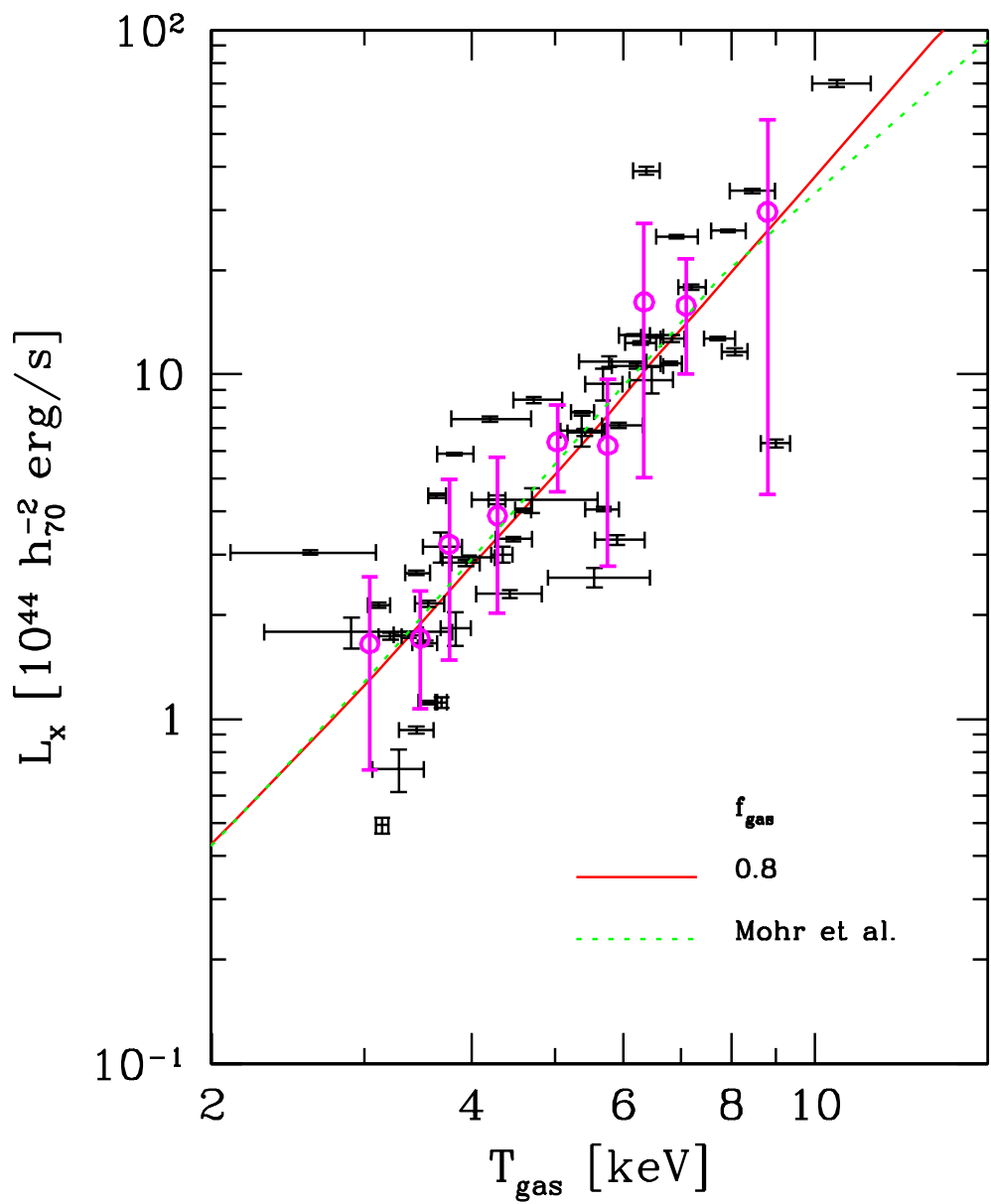}  \figcaption{Best-fit
    $\Lx$-$T$ relations from the parameterized $M$-$T$ relations
    (crosses in Fig.~\ref{fig:f03}). The nine open circles with
    error bars indicate the binned data described in the text. Solid
    and dotted lines correspond to the cases assuming the gas mass
    fraction of $f_{\rm gas}=0.8$ and of
    equation~(\ref{eq:fhotobs500}),  respectively.
    \label{fig:f02}}
\end{figure}

Therefore, we next attempt to find the range of parameters
($p_{\mathit{MT}}$ and $T_{\rm gas, 0}$ in
eq.~[\ref{eq:mtparam}]) which reproduces the observed $\Lx$-$T$
relation. As is clear from Figure~\ref{fig:f01}, the observed data
have intrinsic dispersions. Thus, we first divide the cluster sample in
nine temperature bins so that each bin contains five or six clusters. Then we
compute the mean temperature, and the mean luminosity and the standard
deviation of clusters in each bin. The results are plotted by open
circles with error bars for the luminosity in Figure~\ref{fig:f02}.
Then we perform a $\chi^2$ fit to the binned data.

\begin{figure}[ht]
  \epsscale{0.6} \plotone{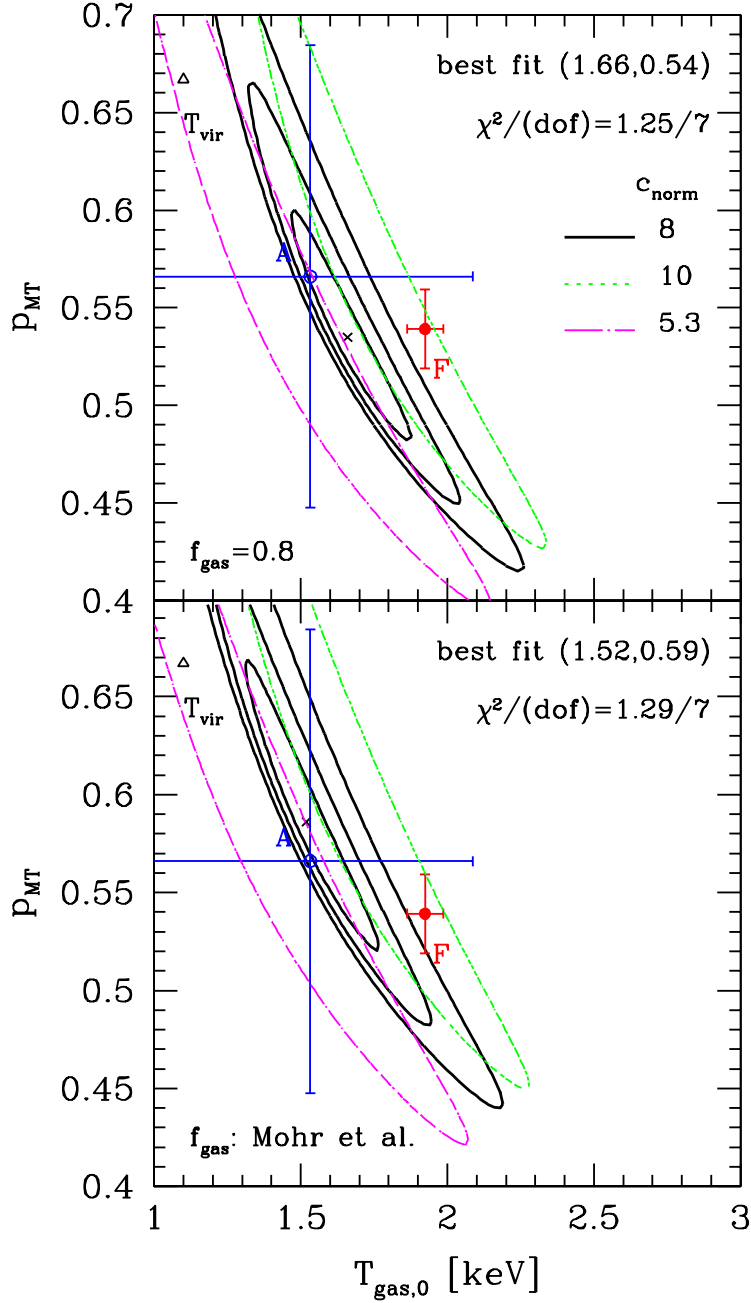} \figcaption{Constraints on
    the parameterized $M$-$T$ relation of clusters from the $\chi^2$
    fit to the binned $\Lx$-$T$ data. Upper and lower panels adopt the
    gas mass fraction of $f_{\rm gas}=0.8$ and of
    eq.~(\ref{eq:fhotobs500}), respectively. The three contour
    curves represent the $1\;\sigma$(68.4\%), $2\;\sigma$(95.6\%), and
    $3\;\sigma$(99.7\%) confidence levels for $c_{\mathrm{norm}}=8$.
    Crosses show the positions of the best-fit values indicated in
    each panel. For comparison, also plotted are the self-similar model
    prediction (\textit{open triangles}), and observational estimates
    by \citeauthor{allen} (\citeyear{allen}; \textit{open circles})
    and \citeauthor{finogu} (\citeyear{finogu}; \textit{filled
      circles}).  Dotted and dot-dashed contours indicate the results
    of $3\;\sigma$(99.7\%) confidence levels for $c_{\mathrm{norm}}=10$
    and 5.3, corresponding to $\pm 1\;\sigma$ deviation from the
    average.  \label{fig:f03}}
\end{figure}

The result is plotted as confidence contours on the $T_{\rm gas,
  0}$-$p_{\mathit{MT}}$ plane in Figure~\ref{fig:f03}. We estimate the 
\textit{relative} confidence levels with respect to the best-fit
values assuming that
\begin{equation}
\Delta\chi^2 \equiv \chi^2(T_{\rm gas, 0}, p_\mathit{MT}) -
\chi^2(T_{\rm gas, 0, min}, p_{\mathit{MT,}{\rm min}})
\end{equation}
follows the $\chi^2$ distribution of the 2 degrees of freedom
\citep[e.g.,][chap. 15]{NR}, where $T_{\rm gas, 0, min}$ and
$p_{\mathit{MT,}{\rm min}}$ are the best-fit parameters that minimize the value
of $\chi^2$. Upper and lower panels correspond to the gas mass
fraction of $f_{\rm gas}=0.8$ and of equation~(\ref{eq:fhotobs500})
from \citet{Mohr99}. The three contour curves represent the 1,
2, and 3$\;\sigma$ confidence levels derived from the value of
$\Delta \chi^2$. Crosses show the positions of the best-fit values
indicated in each panel. For comparison, we plot the self-similar
model prediction, and observational estimates by \citet{allen} and
\citet{finogu} as open triangles, open circles, and filled circles,
respectively. Figure~\ref{fig:f03} rephrases the visual impression
from Figure~\ref{fig:f01} in a more quantitative way; the $M$-$T$
relation of \citet{allen} is inside the 1$\;\sigma$ contour and that of
\citet{finogu} is located just outside the 3$\;\sigma$ confidence level,
and marginally consistent within the large error-bars.  While the
concentration parameter of dark matter halos has a fairly broad
distribution corresponding to $c_{\mathrm{norm}}=8 {+ 2~~ \atop -
  2.7}$ in equation (\ref{eq:c_Bullock}), it does not lead to any
significant difference (compare dotted and dot-dashed lines with solid
lines in Fig.~\ref{fig:f03}). Thus, we fix the proportional constant of
the concentration parameter $c_{\mathrm{norm}}=8$ in what follows.

\begin{figure}[ht]
  \epsscale{0.8} \plotone{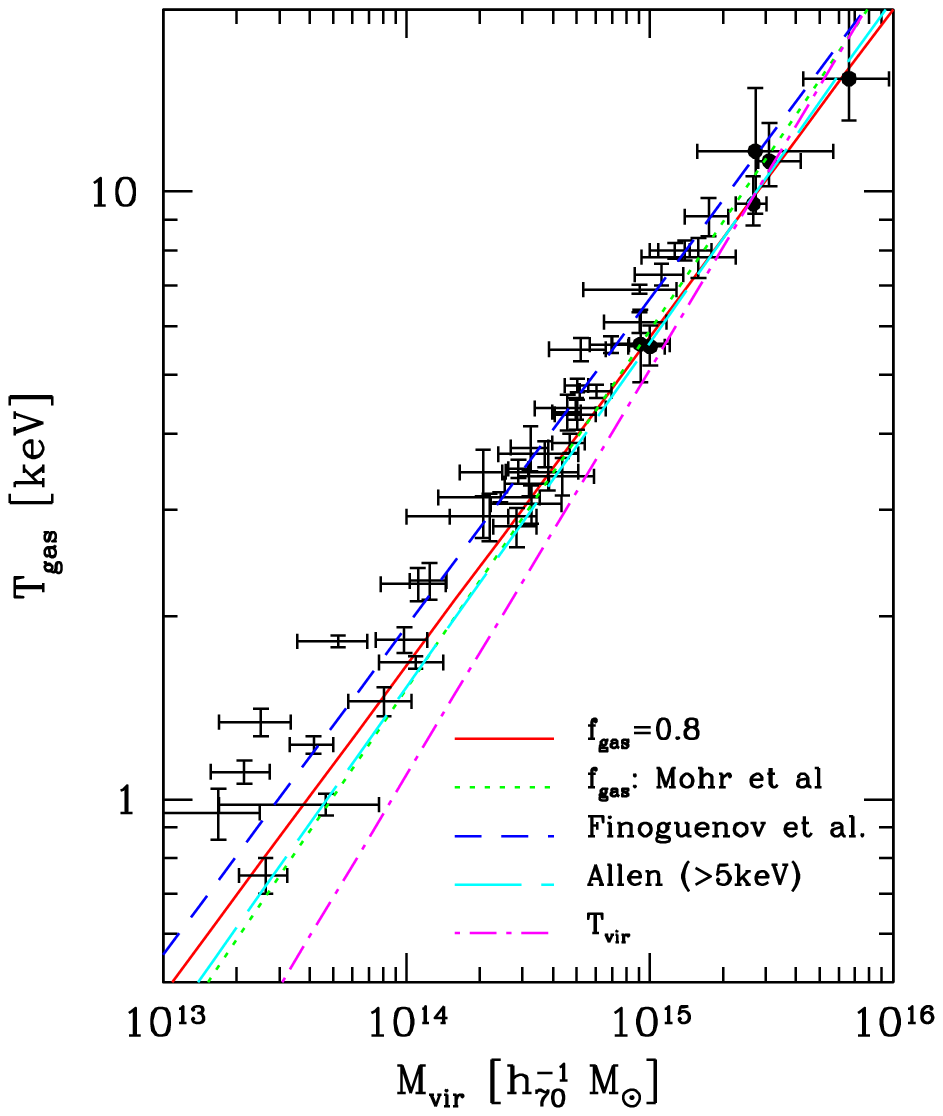}  \figcaption{Predicted
    $M$-$T$ relations compared with observations.  Solid and dotted
    lines indicate the $M$-$T$ relations derived from the observed
    $\Lx$-$T$ relation for $f_{\rm gas}=0.8$ and
    eq.~(\ref{eq:fhotobs500}), respectively (corresponding to the
    crosses in Fig.~\ref{fig:f03}). The observational data
    points and the fits are taken from  \citeauthor{finogu}
    (\citeyear{finogu}; \textit{dots with error bars}) and
    \citeauthor{allen} (\citeyear{allen}; \textit{six filled circles
      with error bars}) after correction for the difference of the mass
    definitions (cf. Appendix A). A simple self-similar model
    prediction with $T_{\rm gas}=T_{\rm vir}$ is plotted in dot-dashed
    line for comparison.
\label{fig:f04}}
\end{figure}

The resulting $\Lx$-$T$ and $M$-$T$ relations for the best-fit
parameters are plotted in Figures~\ref{fig:f02} and \ref{fig:f04}.
While our best-fit models actually reproduce the observed $\Lx$-$T$
relation (the best-fit value of $\chi^2$ per degree of freedom is
shown in each panel of Fig.~\ref{fig:f03}),  they also seem to be in
reasonable agreement with the \textit{observed} $M$-$T$ relation.

\subsection{Constraints from X-ray Temperature Function}

One can also infer the empirical shape of the $M$-$T$ relation from
the requirement that it reproduces the observed XTF of clusters.
Comparing with that from the $\Lx$-$T$ relation, this methodology
provides a fairly (even if not entirely) independent constraint on the
$M$-$T$ relation, mainly in two important aspects:  (1) the model
dependence on the density profile enters only through the flux limit,
$S_{\rm lim}$,  of the observed sample, and (2) the prediction, on the
other hand, is sensitive to the adopted mass function of dark matter
halos, and therefore to the value of $\sigma_8$ in particular
(remember that we fix the other cosmological parameters in the present
analysis).

For this purpose, we again use the 54 X-ray clusters with temperature
higher than 2.5~keV from the sample of \cite{ikebe}. The corresponding
flux limit is $S_{\rm lim} = 2\times 10^{-11}$ erg s$^{-1}$ cm$^{-2}$
in the (0.1--2.4)~keV band, and the total sky coverage is $8.14$ sr.
For definiteness, we adopt the gas mass fraction given by
equation~(\ref{eq:fhotobs500}).  As for the halo mass function, we
adopt both an analytic model by \citet{PS74} and a fitting model to
the numerical simulations by \citet{jenkins},
\begin{eqnarray}
  \frac{dn}{d\ln M_{{\mathrm{vir}}}}
  &=&\frac{\bar{\rho}}{M_{\mathrm{vir}}}f(\sigma)
  \frac{d\ln \sigma^{-1}}{d\ln M_{\mathrm{vir}}},\\
  f(\sigma)&=&0.315\exp\left(-\left|\ln \sigma^{-1}+0.61\right|^{3.8}\right).
\end{eqnarray}
We define the rms variance of linear density field $\sigma$ as
\begin{equation}
  \sigma^{2}(M) =4\pi\int P(k)
  \frac{3}{(kR)^{3}}[\sin(kR)-kR\cos(kR)] k^{2}dk,
\end{equation}
where $R=(3M/4\pi\bar{\rho})^{1/3}$ and $P(k)$ is the linear power
spectrum.  We perform the $\chi^2$ fit to the data with respect to the
three free parameters: $p_{\mathit{MT}}$ and $T_{\rm gas, 0}$ in
equation~(\ref{eq:mtparam}), and $\sigma_8$.

In order to avoid uncertainties in constructing the conventional XTF
data (e.g., the definition of $V_{\rm max}$ as discussed by
\citealt{ikebe}), we directly use the number count of clusters with
the  flux limit $S_{\rm lim}$ per unit solid angle of the sky,
appropriated binned according to their temperatures. In practice, we
divide both the data and our predictions into five temperature bins
(2.5--3.6, 3.6--5.0, 5.0--6.4, 6.4--8.0, and $>$8.0 keV), so that the
errors do not correlate with one another, and perform the $\chi^2$
fit. We assign the Poisson error to the number count in each bin.

\begin{figure}[ht]
  \epsscale{0.6} \plotone{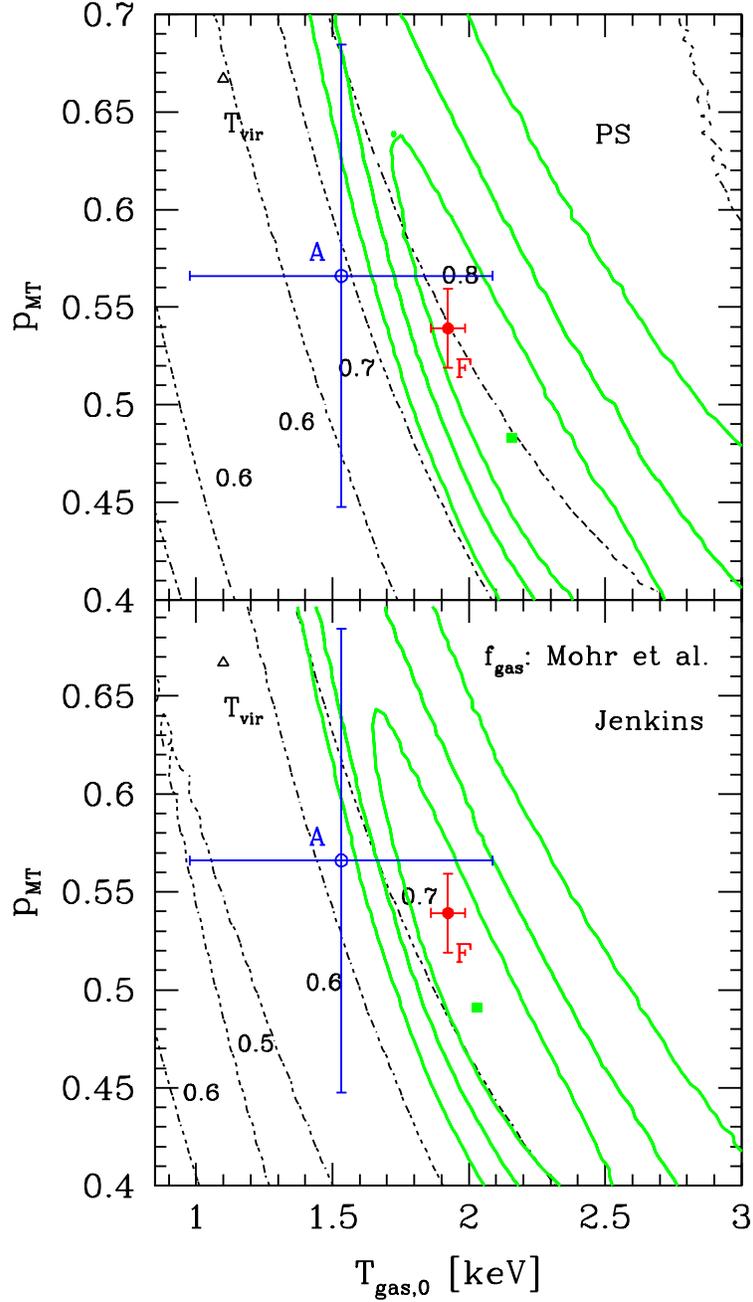}  \figcaption{Constraints on
    the parameterized $M$-$T$ relation of clusters from the $\chi^2$
    fit to the observed XTF. Upper and lower panels adopt the mass
    functions of \citet{PS74} and \citet{jenkins},  respectively. The
    three solid contour curves represent the 1,  2,
    and $3\;\sigma$ confidence levels, with the filled squares
    indicating the points of the highest significance. Dotted contour
    curves show the best-fit values for $\sigma_8$ (which label each
    curve). Symbols are as in
    Fig.~\ref{fig:f03}. \label{fig:f05}}
\end{figure}

The result is plotted as confidence contours on the $T_{\rm gas,
  0}$-$p_\mathit{MT}$ plane in Figure~\ref{fig:f05}. Again we estimate
the \textit{relative} confidence levels with respect to the best-fit
values assuming that
\begin{equation}
\Delta\chi^2 \equiv 
\chi^2(T_{\rm gas, 0}, p_\mathit{MT}, \sigma_{\mathrm{8, local}\
  \mathrm{min}}) - 
\chi^2(T_{\rm gas, 0, min}, p_{\mathit{MT,}\mathrm{min}}, \sigma_{\rm 8, min})
\end{equation}
follows the $\chi^2$ distribution of the 2 degrees of freedom, where
$T_{\rm gas, 0, min}$, $p_{\mathit{MT,}\mathrm{min}}$ and $\sigma_{\rm
  8, min}$ are  the best-fit parameters that minimize the value of
$\chi^2$, and $\sigma_{\mathrm{8, local}\ \mathrm{min}}$ is the value
of $\sigma_8$ minimizing the  $\chi^2$ with given $T_{\rm gas, 0}$,
and $p_\mathit{MT}$. Upper and lower panels of Figure~\ref{fig:f05}
correspond to the mass functions of \citet{PS74} and \citet{jenkins},
respectively. The three solid contour curves represent the 1, 2, and
3$\;\sigma$ confidence levels derived from the $\chi^2$ fit, with the
filled squares indicating the points of the highest significance.
Dotted contours, on the other hand, indicate the best-fit value of
$\sigma_8$ at the given location on the $T_{\rm gas,
  0}$-$p_\mathit{MT}$ plane. The self-similar model prediction, and
observational estimates by \citet{allen} and \citet{finogu}, are
plotted by open triangles, open circles, and filled circles,
respectively.

\begin{figure}[thb]
\begin{center}
  \plotone{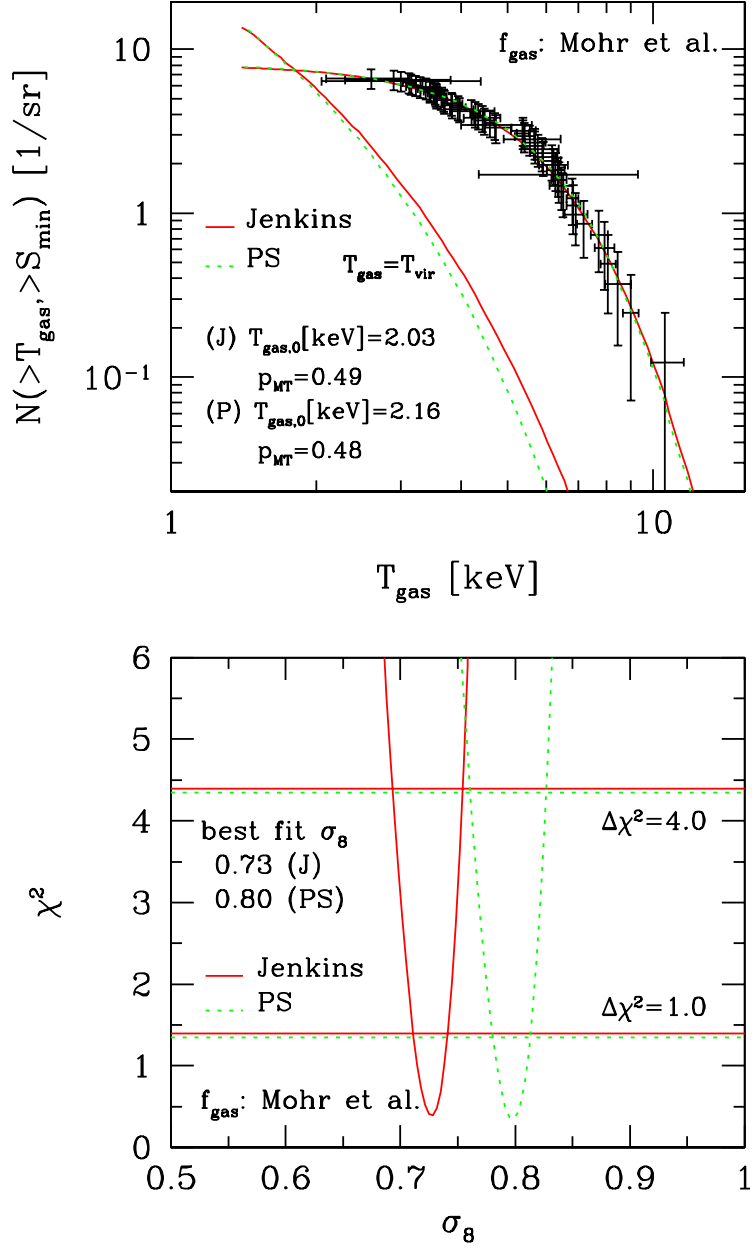}
\end{center}
\figcaption{Best-fit XTF (\textit{upper panel}) from the parameterized
  $M$-$T$ relation compared to the cluster sample of
  \citet{ikebe} and the corresponding $\chi^2$ as a function of
  $\sigma_8$ (\textit{lower panel}).  Solid and dotted curves adopt
  the mass functions of \citet{jenkins} and \citet{PS74},
  respectively. A simple self-similar model prediction with $T_{\rm
    gas}=T_{\rm vir}$ is also plotted just for illustration.
\label{fig:XTF_bestMT}}
\end{figure}

The upper panel in Figure~\ref{fig:XTF_bestMT} illustrates that the
best-fit model reproduces the observed XTF nicely. Again just for
comparison, we plot the self-similar model predictions, which do not
fit the observed XTF at all. Moreover, the lower panel indicates that
the degree of fit is indeed sensitive to the value of $\sigma_8$; the
best-fit value is 0.7--0.8.

Although the above conclusion may seem inconsistent with the previous
claims (\citealt{vl96}; \citealt*{eke96}; \citealt{KS97}) that the
self-similar model does fit the observed XTF with
$\sigma_8=0.9\mbox{--}1.0$, it can be explained by the following
reasons. First, the previous XTF data had much larger errors and thus
still allowed a wider range of theoretical models. Second, the latest
XTF data that we adopted have a systematically smaller amplitude at
$T_{\rm gas} >5\; $keV than the previous ones \citep[see Fig.~6
of][]{ikebe}. Third, the adopted $\Lx$-$T$ relation is different; our
current self-similar model corresponds to $\Lx \propto T_{\rm gas}^2$,
while the previous analyses used (sometimes implicitly) $\Lx \propto
T_{\rm gas}^3$ to construct the ``observed'' XTF data via the
conventional $V_{\rm max}$ method \citep[e.g., ][]{eke96}. Finally,
the recent mass function of \citet{jenkins} predicts more massive
halos than that of \citet{PS74}, which has been widely used in the
previous analyses.

\begin{figure}[thb]
\begin{center}
  \plotone{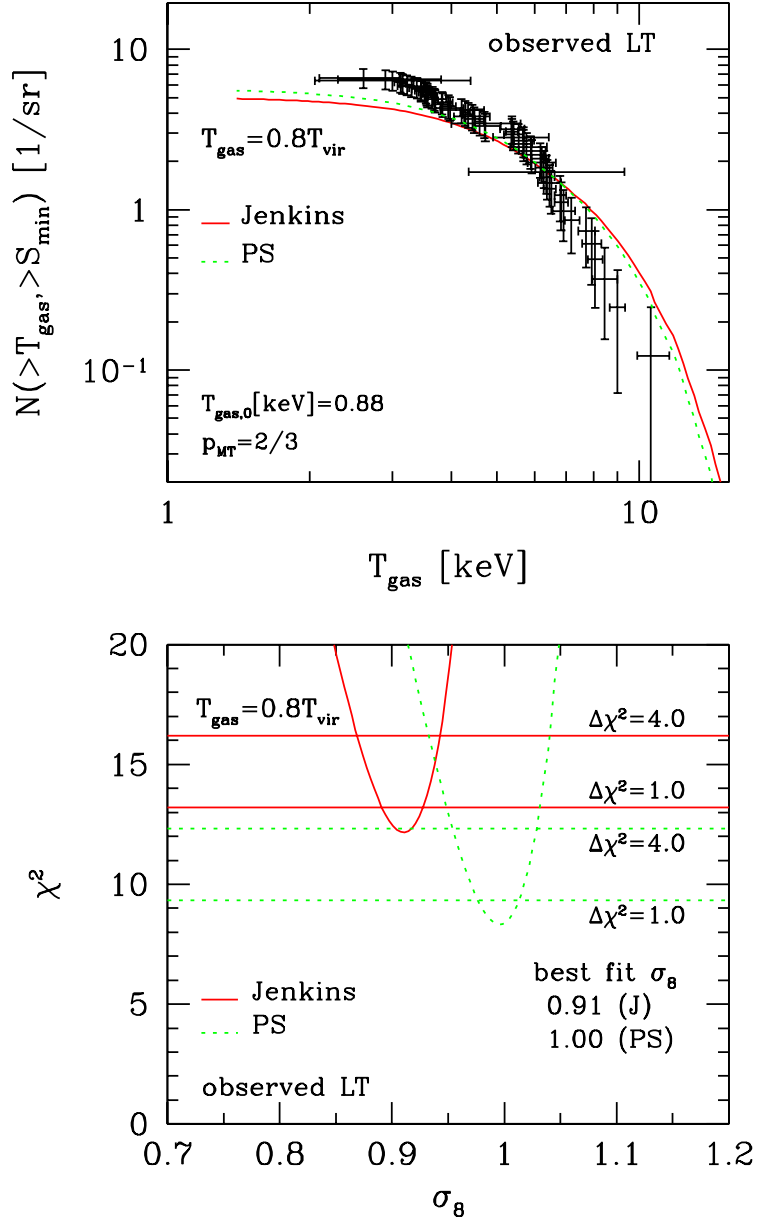}
\end{center}
\figcaption{Same as Fig.~\ref{fig:XTF_bestMT}, but for a self-similar
  model adopting the observed $\Lx$-$T$ relation and a slightly
  smaller proportional factor ($T_{\rm gas}=0.8T_{\rm vir}$).
\label{fig:XTF_SSMT5}}
\end{figure}

Figure~\ref{fig:XTF_SSMT5} demonstrates the above features. For a
direct comparison with the analyses of \citet{KS97}, we adopt $T_{\rm
  gas} = 0.8 T_{\rm vir}$ and the $\Lx$-$T$ relation from equation (3)
of \citet{KS97}, with their fiducial choice for the other parameters.
The goodness of the fit turns out to be significantly degraded
compared to the previous result, mainly because of the systematically
smaller amplitude of the XTF, as well as reduced error bars.
Nevertheless, this methodology fully reproduces the best-fit value of
$\sigma_8=0.9\mbox{--}1.0$, as in the previous one.

In fact, the combination of those effects also explains why the
reanalysis of the cluster abundance by \citet{seljak02} yielded
$\sigma_{8} \sim 0.7$ in the standard $\Lambda$CDM model when adopting
the $M$-$T$ relation of \cite{finogu}. Incidentally, the smaller value
of $\sigma_8$ seems consistent with the recent joint analysis of
cosmic microwave background and large-scale structure \citep{efsta},
which suggests $\sigma_{8} = 0.6\mbox{--}0.7$.

\section{Discussion and conclusions}

We have presented the constraints on the (empirically parameterized)
mass-temperature relation of galaxy clusters from the two different
observational data, the luminosity-temperature relation and the
temperature function.

We summarize in Figure~\ref{fig:f08} our constraints on the $T_{\rm
  gas, 0}$-$p_\mathit{MT}$ plane, adopting the observed gas mass
fraction (eq.~[\ref{eq:fhotobs500}]). The fact that the simple
self-similar evolution model $T_{\rm gas} \propto M_{\rm vir}^{2/3}$
fails to explain the observed $\Lx$-$T$ relation is well known and not
at all new. Rather, it should be noted that the mass-temperature
relation of galaxy clusters is
fairly well constrained by combining the observed $\Lx$-$T$ relation
and XTF. It is encouraging that $T_{\rm gas} =(1.5\sim 2.0)\; {\rm
  keV} (M_{\rm vir}/10^{14}\;h_{70}^{-1}\;M_\odot)^{0.5\sim 0.55}$
barely satisfy the two constraints simultaneously.
This conclusion applies for $1\le \alpha \le 3/2$ as long as the dark
halo density profile is described by equation (\ref{eq:nfw}).

In addition, our analysis implies that the amplitude of the mass
variance $\sigma_8$ in the standard $\Lambda$CDM model should be
0.7--0.8. These values are significantly smaller than the previous
estimates \citep{vl96, eke96, KS97} but in better agreement with more
recent results \citep{seljak02, efsta}.

Since our current analysis has adopted a simple parameterized model
for the mass-temperature relation, the result should be understood by
a physical model of the thermal evolution of the intracluster gas.
For instance, our result may be qualitatively explained by a kind of
phenomenological heating of $T_{\rm gas}(M_{\rm vir}) = T_{\rm
  vir}(M_{\rm vir}) + 1\; $keV. We plan to examine the implications
for the possible heating sources from the derived mass-temperature
relation of galaxy clusters using the Monte-Carlo modeling of merger
trees.

\acknowledgments

We thank an anonymous referee and Alexis Finoguenov for useful
comments.  This research was supported in part by the Grant-in-Aid for
Scientific Research of JSPS (12640231, 14102004). S.S. gratefully
acknowledges support from TMU President's Research Fund for Junior
Scholar Promotion.

\clearpage


\clearpage
\appendix

\section{Mass-radius relation at a given overdensity}
\label{sec:massconvert}

In order to determine the mass of a cluster, one has to specify the
radius from its center. Observationally, this is often set by the
specific fractional overdensity $\Delta_{c}$ with respect to the
critical density of the universe, $\rho_{c}$. Thus the resulting
mass $M_{\Delta_c}$ is written in terms of the corresponding radius
$r_{\Delta_c}$ as
\begin{equation}
 M_{\Delta_{c}}=\frac{4}{3}\pi r_{\Delta_{c}}^{3}
 \Delta_{c}\rho_{c}.
 \label{eq:mdelta}
\end{equation}
Most theoretical studies, on the contrary, usually define the mass of
dark matter halo $M_{\rm vir}$ at its virial radius $r_{\rm vir}$,
adopting the spherical nonlinear collapse model, i.e., $\Delta_{c}
= \Omega_{0}\Delta_{\rm nl}(\Omega_0, \lambda_0)$ (see \citealt{KS96}
for details).

Once the density profile of dark matter halos is specified,
$M_{\Delta_c}$ can be easily translated into $M_{\rm vir}$. For the
profile that we adopted in this paper (eq.~[\ref{eq:nfw}]), mass inside
a radius $r$ is written as
\begin{equation}
 M(r)=4\pi \delta_{c}\bar{\rho}(z) r_{{s}}^{3}
 m(r/r_{{s}}), 
\end{equation}
where $m(x)$ is defined in equation~(\ref{eq:nfw-m}). Thus, if
$\Delta_{\rm c}$ is given, one can compute $r_{\Delta_{c}}$ and
$M_{\Delta_{c}}/M_{\rm vir}$ from the following relation
\begin{equation}
 \frac{M_{\Delta_{c}}}{M_{\rm vir}} =
 \frac{m(r_{\Delta_{c}}/r_{\mathrm{s}})}{m(c)}=
 \frac{r_{\Delta_{c}}^{3}}{r_{\rm vir}^{3}}
 \frac{\Delta_{c}}{\Delta_{\mathrm{nl}}}
 \frac{\rho_{c}}{\bar{\rho}}.
\end{equation}

Figure~\ref{fig:f09} plots $M_{\Delta_{c}}/M_{\rm vir}$
(\textit{upper panels}) and $r_{\Delta_{c}}$ (\textit{lower
  panels})  for several choices of $\Delta_{c}$.

\section{Hot gas mass fraction}

The strategy that we have adopted in the present paper is to determine
the permitted parameter region of the $M$-$T$ relation of clusters
from the $\Lx$-$T$ relation and XTF, \textit{assuming} a specific
model for the hot gas mass fraction $f_{\rm gas}(M_{\rm vir})$
inspired by the observation. In principle, however, we may repeat the
similar procedure and derive the constraints on $f_{\rm gas}(M_{\rm
  vir})$, assuming the observed $M$-$T$ relation instead. Since the
observational uncertainty for the hot gas mass fraction
(eq.~[\ref{eq:fhotobs500}]) that we adopted is fairly large, this
approach is useful in understanding the dependence of our conclusion
on the model for the gas mass fraction.

For this purpose, we also parameterize the gas mass fraction by a
single power law,
\begin{equation}
 f_{\rm gas}(T_{\rm gas})
 =f_{\mathrm{gas}, 0}
 \left(
  \frac{T_{\rm gas}}{1\; \mathrm{keV}}
 \right)^{p_{\scriptscriptstyle\rm gas}}, 
\end{equation}
and perform the $\chi^2$ fit to the $\Lx$-$T$ relation varying the two
free parameters, $f_{\mathrm{gas}, 0}$ and $p_{\rm
  gas}$. We repeat the similar fitting XTF with $f_{\mathrm{gas}, 0}$,
$p_{\rm gas}$ and $\sigma_8$.

The result is plotted in Figure~\ref{fig:f10}, where we use
the observed $M$-$T$ relation of \citet{finogu}. Solid contour curves
represent the 1, 2, and 3$\;\sigma$ confidence levels
derived from the $\chi^2$ fit to the $\Lx$-$T$ relation, while dotted
curves show the 1, 2, and 3$\;\sigma$ confidence levels
from the XTF. Clearly, both constraints again are simultaneously
satisfied for $f_{\rm gas} = (0.5 \pm 0.1)(T_{\rm gas}/1\; {\rm
  keV})^{0.4\pm0.15}$, and in fact agree with the observed gas mass
fraction of \citet{Mohr99}. We also plot the condition that $f_{\rm
  gas}(10\; {\rm keV})=1$ by a  dashed line, so that the hot gas fraction
does not exceed the cosmological average of the baryon fraction
(assuming that $\Omega_{B}=0.04\;h_{70}^{-1}$ and $\Omega_0=0.3$).
If this condition should be satisfied, the acceptable parameter range
almost exactly corresponds to the observed gas mass fraction of
\citet{Mohr99}.

\clearpage

\begin{figure}[ht]
  \epsscale{1.0} \plotone{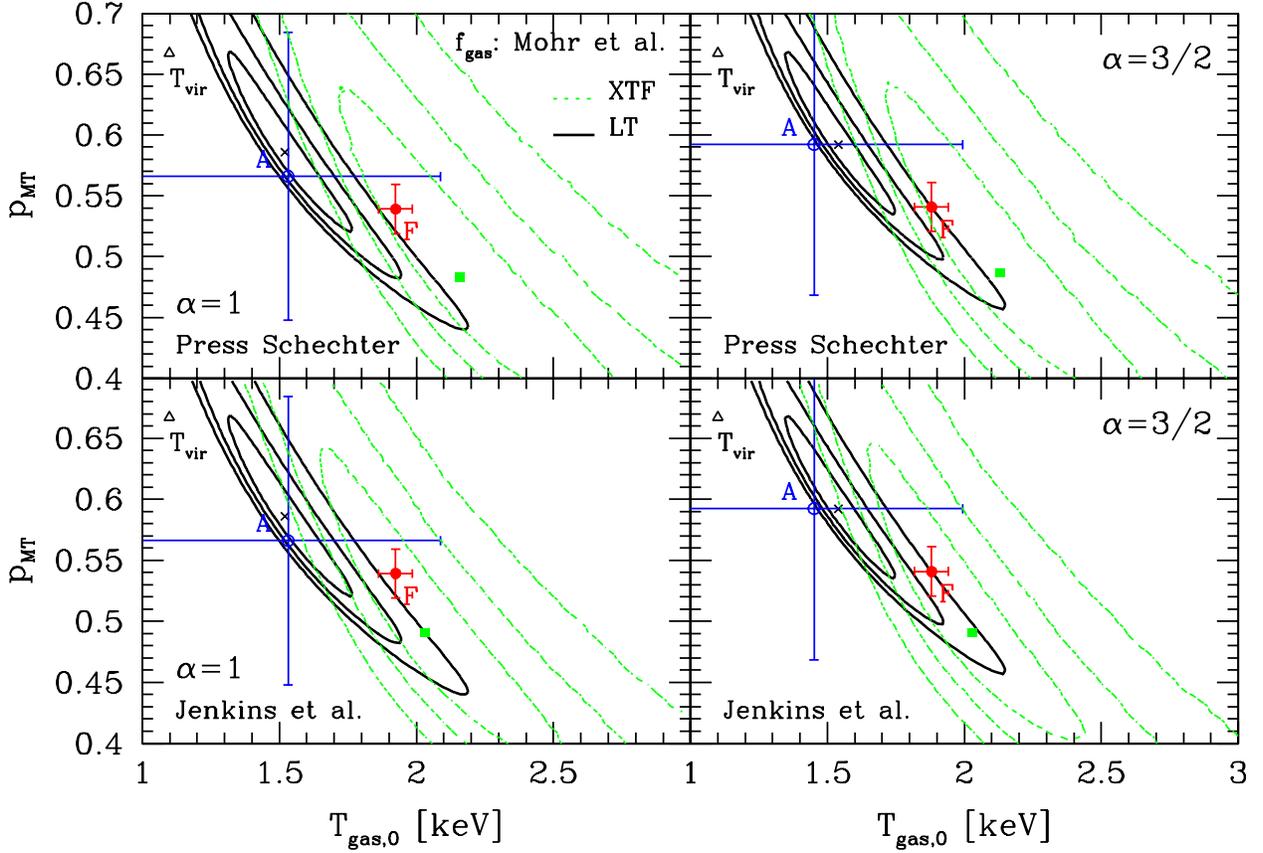}  \figcaption{Joint
    constraints on the parameterized $M$-$T$ relation from the binned
    $\Lx$-$T$ data (\textit{solid curves};  Fig.~\ref{fig:f03})
    and from the XTF (\textit{dotted curves};
    Fig.~\ref{fig:f05}). Upper and lower panels adopt the mass
    functions of \citet{PS74} and \citet{jenkins}, respectively,
    $\alpha=1$ (\textit{Left}) and $\alpha=3/2$ (\textit{Right}).
    Symbols are as in
    Figs.~\ref{fig:f03} and \ref{fig:f05}.
    \label{fig:f08}}
\end{figure}

\begin{figure}[thb]
\begin{center}
  \plotone{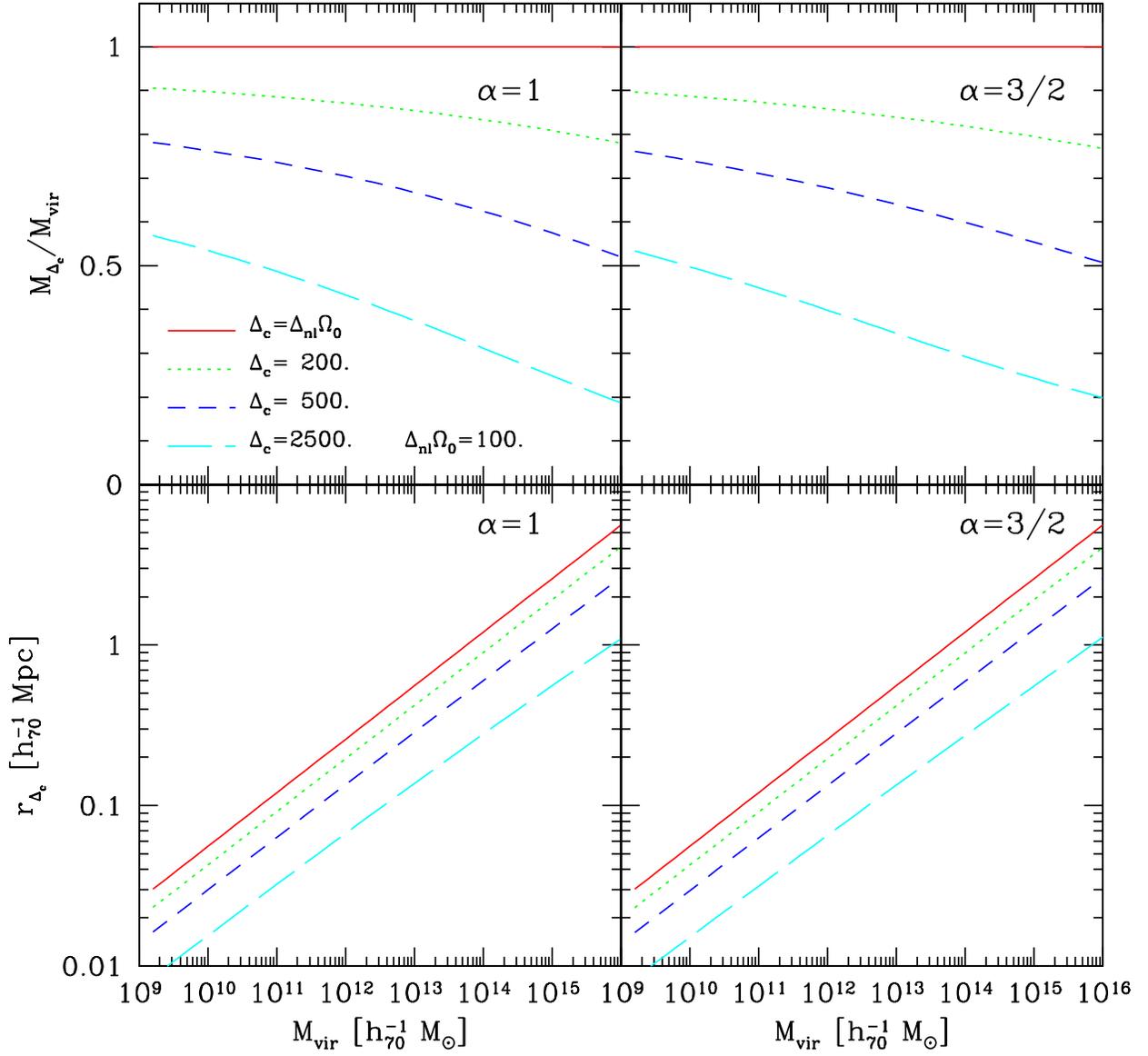}
\end{center}
\figcaption{Ratio, $M_{\Delta_{c}}/M_{\rm vir}$ (\textit{upper
    panels}) and radius, $r_{\Delta_{c}}$ (\textit{lower
    panels}),  for a given overdensity $\Delta_{c}$ as a function
  of $M_{\rm vir}$,  $\alpha=1$ (\textit{left}) and $\alpha=3/2$
  (\textit{right}). Solid,  dotted, short-dashed, and long-dashed
  lines correspond to the overdensities $\Delta_{c} = \Omega_0\,
  \Delta_{\rm nl}$, 200, 500,  and 2500, respectively. The standard
  critical overdensity predicted in the nonlinear spherical collapse
  model $\Omega_0\, \Delta_{\rm nl}$ is $\sim 100$ in our fiducial
  values of the cosmological parameters with $\Omega_0=0.3$ and
  $\lambda_0=0.7$. \label{fig:f09}}
\end{figure}

\begin{figure}[ht]
  \epsscale{0.8} \plotone{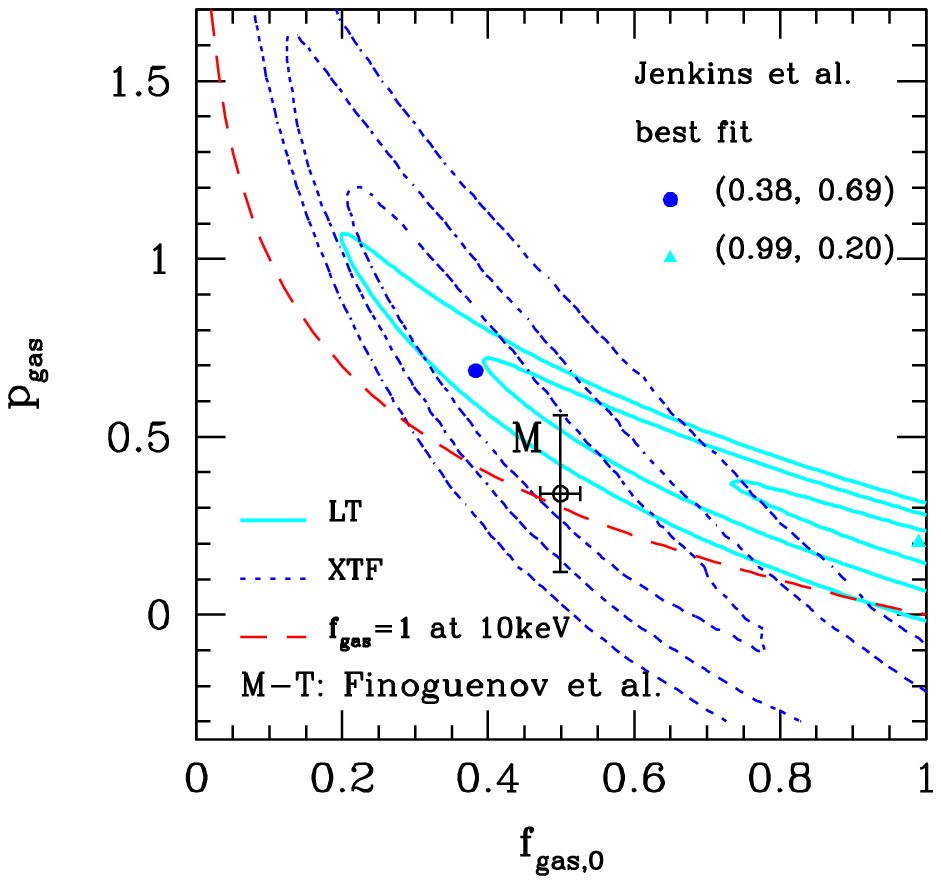}  \figcaption{ Joint
    constraints on the parameterized hot gas mass fraction relation
    from the observed $\Lx$-$T$ relation (\textit{solid curves}) and
    from the XTF (\textit{dotted curves}).  The fit to the
    observational data by \citet{Mohr99} is indicated as an open circle with
    error bars.  Dashed lines indicate the condition that $f_{\rm
      gas}(10\; {\rm keV})=1$.
 \label{fig:f10}}
\end{figure}
\end{document}